\documentclass{revtex4}

\usepackage{graphicx}
\usepackage{dcolumn}
\usepackage{amsmath}
\usepackage{epsfig}
\usepackage{amssymb}

\makeatletter
\def\btt#1{\texttt{\@backslashchar#1}}
\DeclareRobustCommand\bblash{\btt{\@backslashchar}}
\makeatother

\begin{document}

\title{Overview of Quantum Error Prevention and Leakage Elimination}
\author{Mark S. Byrd}
\email{mbyrd@physics.siu.edu}
\affiliation{Physics Department, Southern Illinois University, Carbondale, Illinois
62901-4401}
\author{Lian-Ao Wu}
\email{lwu@chem.utoronto.ca}
\author{Daniel A. Lidar}
\email{dlidar@chem.utoronto.ca}
\affiliation{Chemical Physics Theory Group, Chemistry Department, University of Toronto,
80 St. George Street, Toronto, Ontario M5S 3H6, Canada}
\date{\today}

\begin{abstract}
Quantum error prevention strategies will be required to produce a scalable
quantum computing device and are of central importance in this regard.
Progress in this area has been quite rapid in the past few years. In order
to provide an overview of the achievements in this area, we discuss the
three major classes of error prevention strategies, the abilities of these
methods and the shortcomings. We then discuss the combinations of these
strategies which have recently been proposed in the literature. Finally we
present recent results in reducing errors on encoded subspaces using
decoupling controls. We show how to generally remove mixing of an encoded
subspace with external states (termed leakage errors) using decoupling
controls. Such controls are known as ``leakage elimination operations'' or
``LEOs.''
\end{abstract}

\maketitle



\section{Introduction}

Preventing errors in quantum information is an important part of quantum
information theory and a central goal in quantum computing. Since efficient
algorithms make use of many particle quantum states which are very fragile,
this will be a key component of any working quantum computing device.

An idealistic goal would be the noiseless evolution of the quantum system.
(We will take ``noise" to mean both unitary errors and decoherence in a
quantum system throughout this article and will specify if and when the need
arises.) However, it is clear that no system is noiseless since it will
always interact with an environment and we cannot implement any operation
perfectly. Thus after isolating a system to the best of our ability, we
should aim for the realistic goals of the identification and correction of
errors when they occur and/or avoiding noises when possible and/or
suppressing noise in the system. To each of these tasks there corresponds an
error prevention strategy developed for the specific purpose; quantum error
correcting codes (QECCs), decoherence-free or noiseless subsystems (DFSs)
and \textquotedblleft bang-bang\textquotedblright\ decoupling controls (BB).
All three of these classes of error prevention have limitations. Therefore
the choice of error prevention protocol depends on the system. Hybrid
strategies appear to be required for near-future experiments in which qubits
are available in a limited supply.

In the first part of this article we give an overview/review of the three
error prevention strategies. This is designed to be a resource for novices
as well as experts which conveys the ideas, objectives and several key
references for quantum error prevention schemes. This includes an
introduction to, and the range of applicability of, the three quantum error
prevention classes. We then review some proposals for the combinations of
these methods. Finally, we discuss recent results concerning the elimination
of errors which serve to destroy the effectiveness of encoded qubits.


\section{Quantum Error Correction Strategies}


\subsection{Quantum Error Correcting Codes}

Very generally, we may describe a quantum error correcting code as a set of
states which can be used to store information in a way that errors are able
to be detected and corrected during a quantum information processing task.
As with classical error correcting codes, the code is a repetition or
redundancy code where information is stored in a state within a subspace.  
Errors are correctable as long as they map orthogonal states to orthogonal
states. The first to show how to implement a quantum error correcting code
(QECC) were Shor \cite{Shor:95} and Steane \cite{Steane} who proved the
in-principle possibility of correcting errors in quantum computing devices.
Shor's code uses nine physical qubits to encode one logical qubit and thus
stores one qubit of information reliably. It protects the logical qubit
against single independent errors on the physical qubits and is denoted
[9,1,3]. The first entry is the number of physical qubits, the second the
number of logical qubits and the third is the distance. ($d=2t+1$ where $t$
is the number of errors which the code can protect against.) Subsequently
several methods were discussed for the construction of $[n,k,d]$ codes \cite%
{Calderbank:96,Steane,Gottesman:97,Gottesman:97b}. (A lucid account of the
precise requirements is given in \cite{Knill:97b}.)

Several authors investigated quantum error correcting codes, showing that
there are large classes of such codes. Two especially important classes are
the CSS codes (Calderbank and Shor \cite{Calderbank:96} and Steane \cite%
{Steane}) and their generalization, the stabilizer codes \cite%
{Gottesman:97,Gottesman:97b}. In addition to the descriptions of the classes
of codes which protect against different types of errors, a bound was
obtained, called the quantum Hamming bound, which describes the smallest set
of states needed to protect against a given set of errors and defines
efficient codes \cite{Ekert:96a,Knill:97b}. This sets the limit of five for
the number of physical qubits needed to protect one logical qubit against
arbitrary single independent errors on the physical qubits. When errors are
not independent or when gating errors are present, the number of physical
qubits required to encode one logical qubit grows dramatically. In addition,
storage and gating errors must be below a certain threshold for this scheme
to work reliably. (See \cite{Preskill:99} and references therein for the
threshold as well as fault-tolerant recovery requirements.) These
constraints imply QECCs are the least qubit intensive when gating errors are
low and the errors are truly independent.

In the near future (perhaps before 2010), we expect to have fewer than 50
physical qubits available in quantum computing experiments. Therefore,
physical qubits will be a scarce resource. An encoding into a QECC would
demand that 50 logical qubits are reduced to, at most, ten (neglecting
ancilla qubits which are used for fault-tolerant recovery). Ten qubits can
be created for use in NMR experiments at this time and proof-of-principle
experiments have already been performed to exemplify the use of QECCs \cite%
{Knill:01}. We therefore seek error prevention methods which provide a
higher ratio of the number of logical qubits to physical qubits in order to
investigate a wider range of scaling issues and algorithms. This is, in
large part, the motivation for hybrid error prevention techniques, discussed
below, which are sought for use in experiments which will be performed in
the next ten years.


\subsection{Decoherence-free Subspaces and Noiseless Subsystems}

A decoherence-free subspace and its generalization, a noiseless or
decoherence-free subsystem (DFS) , is a state or set of states which is not
vulnerable to decoherence \cite%
{Zanardi:97c,Duan:98,Lidar:PRL98,Knill:99a,Kempe:00,Lidar:00a}. (For a
recent review see \cite{Lidar/Whaley:03}.) In this case one takes advantage
of a symmetry in a system-bath interaction in order to store information in
a DFS which is invariant under the action of the interaction Hamiltonian.
Under appropriate circumstances one can expect such a symmetry to exist.
However, identifying a useful symmetry and taking advantage of it can be
very difficult. One must 1) identify the symmetry, 2) find the states which
are invariant to the interaction and 3) construct, if possible, operations
on the system which will serve as a universal set of gating operations 
\textsl{and} preserve the necessary symmetries. Although this may seem a
daunting task, DFSs have been found which satisfy all of these requirements.

DFSs have shown promise in several experiments and have been observed to
reduce noise in others \cite%
{Kwiat:00,Kielpinski:01,Viola/etal:01,Fortunato/etal:02}, including
computation in a DFS \cite{Ollerenshaw:03,Mohseni:03}. DFSs have also led to
the concept of encoded universality (find subspaces in which universal
quantum computing can be performed even when it is not possible to perform
universal computing on the whole space) \cite%
{Bacon:99a,Kempe:00,Bacon:Sydney,DiVincenzo:00a,Levy:01,Benjamin:01,Wu/Lidar:01a,Lidar/Wu:01,Kempe:01d,Kempe:01e,Lidar/Wu/Blais:02}%
.

For near future experiments, DFSs have advantages over quantum error
correcting codes since the number of physical qubits required to encode one
logical qubit is typically lower, and they do not require repeated
identification and correction of errors. Once the qubits are encoded, they
evolve noiselessly. The disadvantage is the difficulty in identifying the
symmetry and exploiting it. Nevertheless, even when a symmetry cannot be
found, it can be actively generated, a procedure known as encoded decoupling 
\cite%
{Lidar/Wu:01,Wu/Lidar:cdfs,Byrd/Lidar:ss,Wu/Lidar:01a,Wu/Lidar:cdfs,Wu/etal:02,Lidar/Wu:02,Lidar/Wu:03}%
.


\subsection{Dynamical Decoupling}

Bang-bang decoupling (BB) operations can be traced back to the decoupling
operations used in NMR experiments \cite{Haeberlen/Waugh,Ernst:book}. In the
simplest case, one lets the Hamiltonian evolve for a time $t$, then changes
the open-system evolution by acting only on the system, and lets it evolve
for a second time $t^{\prime }$. This produces an effective evolution after
a total time $t+t^{\prime }$. If the time evolution (the Hamiltonian) can be
inverted for a time $t^{\prime }=t$, then the two evolutions will cancel,
producing zero net evolution after time $2t$. They may therefore be used to
eliminate Hamiltonian evolutions.

The first uses for the purpose of general noise reduction in quantum
computing systems are found in \cite{Viola:98,Ban}. These showed that within
a spin-boson model 
strong, fast operations can be used to eliminate the interaction with the
environment which causes dephasing of a qubit. However, BB can be viewed
more generally as a symmetrization, or averaging, technique which is more
general than simply inverting the time evolution directly \cite%
{Zanardi:98b,Viola:99,Byrd/Lidar:01,Viola:01a}. Several extensions of this
method have since been given, including conditions for computing in the
presence of the decoupling controls \cite{Viola:99a} and for using empirical
data to determine an appropriate set when computing or not \cite%
{Byrd/Lidar:ebb}.

The motivation for studying this technique more thoroughly is clear: \textit{%
dynamical decoupling controls do not require extra qubits} for the
reduction/elimination of noise and decoherence in the quantum system. This
is a major advantage since, as stated above, physical qubits will be a
scarce resource in near future experiments. The limitations of dynamical
decoupling is that they assume that the control operations are strong and
fast \cite%
{Viola:98,Ban,Duan:98e,Viola:99,Zanardi:98b,Vitali:99,Byrd/Lidar:ebb,Viola:99a,Viola:00a,Vitali:01,Byrd/Lidar:01,Cory:00,Agarwal:01,Uch:02,Uch:03}. 
In fact there is a strong connection to the quantum Zeno effect (QZE) 
\cite{Facchi/etal:04}.  There are
certainly systems for which neither of these assumptions is difficult to
satisfy and the strong assumption is less stringent \cite{strongnote}.
However, the fast assumption can be quite difficult to satisfy, and if it is
not satisfied decoherence can be accelerated rather than suppressed, as in
the inverse QZE \cite{Facchi/etal:04}. Roughly speaking, we require a complete
set of control operations to be implemented within the correlation time of
the bath \cite{Viola:98,Viola:99}. This is due to the fact that one aims to
eliminate the system-bath interaction before the information is
irretrievably lost, quite opposite of a Markovian assumption. There is a
notable exception however; in the case of $1/f$ noise BB has been shown not
to crucially depend on the high-frequency bath cut-off (which is the inverse
of the correlation time of the bath) \cite{Shiokawa/Lidar:02,Faoro:03}.  
This implies that the fast requirement is not difficult to satisfy in some
important cases \cite{Lidar/Wu:02}. The strong requirement can also be
relaxed if certain conditions apply \cite{Viola:02}.

Soon after the initial research into BB, several authors sought to combine
BB with DFSs \cite%
{Zanardi:98b,Viola:00a,Wu/Lidar:cdfs,Byrd/Lidar:ss,Byrd/Lidar:ebb,Viola:01a}%
. This is, in part, the subject of the next section: combining error
prevention techniques.


\section{Combining Methods of Error Prevention}

\label{sec:combining}

Given the three different error prevention methods described above several
possibilities for combinations exist. To be specific, we could combine

\begin{enumerate}
\item DFSs and QECCs, \label{DFSQECC}

\item QECCs and BB, \label{QECCDD}

\item DFSs and BB, \label{DFSDD}

\item QECCs, DFSs and BB. \label{QECCDFSDD}
\end{enumerate}

At this time, all of these combinations have been explored to varying
degrees in the literature. The first combination, DFSs and QECCs, is
described in \cite{Lidar:PRL99,Lidar:00b}. In the case \cite{Lidar:PRL99} a
perturbative independent error on a DFS structure may be detected and
corrected. In the case \cite{Lidar:00b} computation takes the encoded
information out of and then into the DFS, and errors that occur along this
trajectory may be corrected by a compatible QECC. The subject was also
studied for \textquotedblleft detected-jump error
correcting\textquotedblright\ (detecting a spontaneously emitted excitation)
in \cite{Alber/etal:01,Alber/etal:02a,Alber/etal:02b,Khodjasteh/Lidar:02}. 
The idea is to let the DFS encoding protect against the non-unitary,
conditional evolution that arises in the quantum trajectories picture, and
let the QECC correct the errors that arise during quantum jumps in the same
picture. The second combination, QECCs and BB, can be achieved in at least
two ways. First, one can use BB on each physical qubit in the code to reduce
noise on that particular independent qubit. One could also use BB on the
logical qubits using the stabilizer formalism to determine the appropriate
decoupling sequence \cite{Byrd/Lidar:ss,Byrd/Lidar:pqe02}. 
A third possibility is to use BB to suppress the conditional evolution and
use QECC\ to correct quantum jumps, in the quantum trajectories picture 
\cite{Khodjasteh/Lidar:03}.  
This combination is interesting in the sense that it uses a
minimal QECC ($n+1$ physical qubits per $n$ logical qubits) and applies BB\
in the Markovian regime. The third combination, DFSs and BB, has been the
most thoroughly explored. There are several different motivations for this
but the primary objective is to produce an effective evolution which is
compatible with a DFS \cite%
{Zanardi:98b,Viola:00a,Wu/Lidar:cdfs,Byrd/Lidar:ss,Viola:01a,Byrd/Lidar:ebb,Byrd/Lidar:pqe02,Lidar/Wu:02,Wu/etal:02,Zanardi:99a}%
. In all of these scenarios, the demands on the physical system have been
drastically reduced by requiring only that the Hamiltonian be \textit{%
modified} by BB in order to produce an interaction Hamiltonian which is
compatible with an encoding method. This is in contrast to the original
decoupling proposals which required that the interaction Hamiltonian be 
\textit{eliminated}.

In principle, BB can be combined with any encoding \cite%
{Byrd/Lidar:ss,Byrd/Lidar:ebb,Zanardi:98b,Zanardi:99a}. The codewords could
then be DFS codewords, QECC codewords or any combination thereof. However,
more specific results exist for the combination of all three methods in
order to actively produce the conditions for a DFS and for correction of the
departure from the symmetry required for a DFS using QECC techniques \cite%
{Lidar:00b,Wu/Lidar:cdfs}. For the next few sections we discuss methods of
eliminating errors on a predefined qubit (encoded or physical) using BB.


\section{Eliminating Leakage}

We now discuss a specific combination of encoding and BB methods called
\textquotedblleft leakage elimination.\textquotedblright\ We begin by
reviewing previous results, then presenting some new results and finally we
provide examples of physical systems where such techniques are useful.

An ideal qubit is a two-level system consisting of a pair of orthonormal
quantum states. However, this idealization neglects other levels which are
typically present and can mix with those defining the qubit. This mixing is
what we will refer to as \textquotedblleft leakage\textquotedblright .
Leakage may be the result of the application of logical operations, or
induced by system-bath coupling. In the former case, a rather general
solution was proposed in \cite{Tian:00}. In the next few sections we will be
interested in decoherence-induced leakage. The logical qubits of codes, as
well as physical qubits, can undergo leakage errors, which are particularly
serious: by mixing states from within the code and outside the code space,
leakage completely invalidates the encoding. A simple procedure to detect
and correct leakage, which can be incorporated into a fault-tolerant QECC
circuit, was given in \cite{Preskill:97a}. This scheme is, however, not
necessarily compatible with all encodings \cite{Kempe:01d}. Here we present
a \emph{universal}, BB solution to leakage elimination in the limit of fast
and strong \textquotedblleft bang-bang\textquotedblright\ (BB)\ pulses \cite%
{Viola:98,Vitali:99,Zanardi:99,Viola:99}.


\subsection{Leakage Elimination Operators}

Suppose that several multilevel systems are used to encode $2^K$ logical
states representing $K$ qubits (with the appropriate tensor product
structure). Let us arrange the basis states $\{\left\vert k\right\rangle
\}_{k=0}^{N}$, $N=2^K$ of $\mathcal{H}_{N}$ so that $\left\vert
0\right\rangle_i$ and $\left\vert 1\right\rangle_i$ ($i\in K$) represent the
physical or encoded (logical) qubit states. The code subspace will be
denoted $C$ and its orthogonal complement $C^\perp$. We can classify all
system operators as follows: 
\begin{equation}
E=\left( 
\begin{array}{cc}
\sigma_L & 0 \\ 
0 & 0%
\end{array}
\right) \quad E^{\bot }=\left( 
\begin{array}{cc}
0 & 0 \\ 
0 & \sigma_L^\perp%
\end{array}
\right) \quad L=\left( 
\begin{array}{cc}
0 & D \\ 
F & 0%
\end{array}
\right) ,  \label{eq:EEL}
\end{equation}
where $\sigma_L$ and $\sigma_L^\perp$ correspond to operations on the
logical states and the orthogonal subspace respectively. Operators of type $%
E $ produce logical operations, i.e., they act entirely within the code
subspace. $E^\perp$ operators, on the other hand, have no effect on the code
as they act entirely outside the qubit subspace.  
Finally, $L$ represents the leakage operators, with $D,F$ off-diagonal
blocks which have the effect of creating superpositions between states
within a code and outside of the code subspace. These algebraic elements
correspond to the leakage from, or to, the logically encoded subspace.

Let us now recall the \textquotedblleft parity-kick\textquotedblright\
scheme \cite{Viola:98,Vitali:99}, which is a special case of BB. The total
system-bath Hamiltonian can be written as $H_{SB}=H_{C}+H_{\perp }+H_{L},$
where $H_{C}$ ($H^{\perp },H_{L}$) is a linear combination of elements of
the set $E$ ($E^{\perp },L$) tensored with bath operators. Now consider a 
\emph{leakage-elimination operator} (LEO) 
\begin{equation}
R_{L}=e^{i\phi }\left( 
\begin{array}{cc}
-I & 0 \\ 
0 & I%
\end{array}%
\right) ,  \label{eq:R_Lmat}
\end{equation}%
where the blocks have the same dimensions as in Eq.~(\ref{eq:EEL}) and $\exp
(i\phi )$ is an overall phase factor. This operator anticommutes with the
leakage operators: $\{R_{L},L\}=0$, while $[R_{L},E]=[R_{L},E^{\bot }]=0$.
It is an LEO since it follows that the following (parity-kick) sequence
eliminates the leakage errors: 
\begin{equation}
\lim_{n\rightarrow \infty }(e^{-iH_{SB}t/n}R_{L}^{\dagger
}e^{-iH_{SB}t/n}R_{L})^{n}=e^{-iH_{E}t}e^{-iH^{\bot }t}  \label{e1}
\end{equation}%
To physically implement this, in practice one takes $n=1$ and makes the time 
$t$ very small compared to the bath correlation time \cite%
{Viola:98,Vitali:99}. Eq.~(\ref{e1}) then holds to order $t^{2}$, and
implies that one intersperses periods of free evolution for time $t$ with $%
R_{L}$, $R_{L}^{\dagger }$ pulses which are so strong that $H_{SB}$ is
negligible during the BB pulses. The term $e^{-iH^{\bot }t}$ in Eq.~(\ref{e1}%
) has no effect on the qubit subspace. The term $e^{-iH_{E}t}$ may result in
logical errors, which will have to be treated by other methods, e.g.,
concatenation with a QECC \cite{Preskill:97a,Knill:98,Lidar:PRL99}, or
additional pulses \cite{Viola:99,Byrd/Lidar:01,Byrd/Lidar:ss}. Therefore, in
order to eliminate \emph{leakage}, we seek an LEO for a given encoding,
which is obtainable from a controllable system Hamiltonian $H_{S}$ acting
for a time $\tau $, i.e., $R_{L}=\exp (-iH_{S}\tau )$.


\subsection{Generalized LEO}

The leakage operator given in \cite{Wu/etal:02} was termed canonical if the
corresponding Hamiltonian was also a projection operator onto the code space 
$C$. The physically available logical operations may or may not be canonical
in this sense. Here we show that we may relax this restriction and that one
may obtain an LEO that need not also be a projective operation. In Section %
\ref{sec:4qubitDFS} we will give an explicit example of such an operator.

In (\cite{Zanardi:98b,Wu/etal:02}) it was shown that an LEO $R_L$ may be
obtained through the exponentiation of a Hamiltonian in the following form 
\begin{equation}  \label{eq:R_L^E}
R_L = \exp(\pm i\pi \sigma_L P),
\end{equation}
where $P$ is a projection and $\sigma_L$ any operation such that $%
\sigma_L=\sigma_L^\dagger$ and $\sigma_L^2=1$, e.g., a logical operation.
Note that the logical operations very often are already projective in the
sense that they operate only on the code space. Examples will be given below.

However, not all LEOs have such a form and we now give a more general
characterization of an LEO. Let the Hamiltonian for an LEO be given by 
\begin{equation}
H = \left( 
\begin{array}{cc}
H_1 & 0 \\ 
0 & H_2%
\end{array}
\right),
\end{equation}
where $H_1$ acts on the code subspace and $H_2$ on the orthogonal
complement. If $H_1$ is diagonal with even (odd) integers as the diagonal
elements and $H_2$ is diagonal with odd (even) integers as the diagonal
elements, then one may write the LEO as 
\begin{equation}
R_L = U \exp(- i \pi H) U^\dagger
\end{equation}
where $U = U_1 \oplus U_2$ is a direct sum (block diagonal). In this case $H$
is not projective since it may have non-zero eigenvalues when acting on the
subspace orthogonal to the code. The effective LEO, however, is unchanged,
i.e., the form Eq.~(\ref{eq:R_Lmat}) is obtained which again produces and
thus eliminates leakage errors as desired. Such is the case for the
four-qubit DFS example in Section~\ref{sec:4qubitDFS}.

We should note at this point that this can immediately be generalized to an
arbitrary number of qubits (physical or logical qubits) \cite{Wu/etal:02}.
We will review several physical examples of leakage elimination from Ref.~%
\cite{Wu/etal:02} in the next section.


\subsection{Examples of Leakage Elimination in Physical Systems}

\textit{Example 1.---} As a simple first example, consider physical qubits
(without encoding), such as electrons on liquid helium \cite{Platzman:99},
or an electron-spin qubit in quantum dots \cite%
{Loss:98,Levy:01a,Imamoglu:99,Pazy:01}, or a nuclear-spin qubit in donor
atoms in silicon \cite{Kane:98,Vrijen:00}. In those cases, a potential well
at each site traps one fermion. Usually, the ground and first excited state
are taken as a qubit for a given site: $|k\rangle =c_{k}^{\dagger }|\mathrm{%
vac} \rangle $, where $c_{k}^{\dagger }$ is a fermionic creation operator
for level $k=0,1$. Let $n_{k}=c_{k}^{\dagger }c_{k}$ be the fermion number
operator. The logical operations for this qubit are $E=\{X=c_{0}^{%
\dagger}c_{1}+c_{0}^{\dagger }c_{1},
Y=i(c_{1}^{\dagger}c_{0}-c_{0}^{\dagger}c_{1}),Z=n_{0}-n_{1}\}$ whose
elements satisfy $su(2)$ commutation relations. In this case, a general
linear Hamiltonian which includes hopping terms, $H_{SB}=%
\sum_{k,l=0}^{N-1}a_{kl}c_{k}^{\dagger }c_{l}$, where $a_{kl}$ includes
parameters and bath operators, and $k,l$ denote all electron states, can
leak the qubit states $k=0,1$ into any of the other states. Using
parity-kicks, we can eliminate this leakage in terms of the LEO: $R_{L}^{%
\mathrm{id}(1)}=\exp [\pm i\pi (n_{0}+n_{1})]$. This LEO is implemented
simply by controlling on-site energies.

\textit{Example 2}.--- We can also treat bosonic systems, such as the linear
optical QC proposal \cite{Knill:00}. In this case, a qubit is encoded into
two modes. The first qubit has states $|0\rangle _{1}=b_{1}^{\dagger }| 
\mathrm{vac}\rangle $ and $|1\rangle _{1}=b_{2}^{\dagger }|\mathrm{vac}
\rangle $, and the second qubit is $|0\rangle _{2}=b_{3}^{\dagger }|\mathrm{%
\ vac}\rangle $ and $|1\rangle _{2}=b_{4}^{\dagger }|\mathrm{vac}\rangle $,
where $b_{i}^{\dagger }$ are bosonic creation operators. Encoded two-qubit
states are $|00\rangle =b_{1}^{\dagger }b_{3}^{\dagger }|\mathrm{vac}\rangle
,$ $|01\rangle =b_{1}^{\dagger }b_{4}^{\dagger }|\mathrm{vac}\rangle ,\
|10\rangle =b_{2}^{\dagger }b_{3}^{\dagger }|\mathrm{vac}\rangle $ and $%
|11\rangle =b_{2}^{\dagger }b_{4}^{\dagger }|\mathrm{vac}\rangle .$ But the
linear optical Hamiltonian $H=\sum_{k,l=1}^{4}a_{kl}b_{k}^{\dagger }b_{l}$,
contains beam-splitter terms like $b_{1}^{\dagger }b_{3}$ and $%
b_{2}^{\dagger }b_{3}$, which can cause leakage into states such as $%
b_{1}^{\dagger }b_{2}^{\dagger }|\mathrm{vac}\rangle $ or $b_{1}^{\dagger
2}| \mathrm{vac}\rangle $. By using the LEO $R_{L}^{\mathrm{id}(1)}=\exp %
\left[ \pm i\pi (b_{1}^{\dagger }b_{1}+b_{2}^{\dagger }b_{2})\right] $, we
can eliminate the leakage terms. This LEO can be implemented simply using a
linear optical phase shifter. However, generalizing this LEO\ to multiple
encoded qubits requires a photon-photon interaction, which is not readily
available.

\textit{Example 3}.--- A substantial number of promising solid-state QC\
proposals, e.g. \cite%
{Platzman:99,Loss:98,Levy:01a,Imamoglu:99,Pazy:01,Kane:98,Vrijen:00,Mozyrsky:01}%
, are governed by effective isotropic and anisotropic exchange interactions,
which quite generally, can be written as 
\begin{equation}
H_{\mathrm{ex}}=%
\sum_{i<j}J_{ij}^{x}X_{i}X_{j}+J_{ij}^{y}Y_{i}Y_{j}+J_{ij}^{z}Z_{i}Z_{j},
\label{eq:Hex}
\end{equation}%
where $X_{i}$ is the Pauli $\sigma _{x}$ matrix on the $i$th qubit, etc. The
encoding $\left\vert 0\right\rangle _{L}=|01\rangle $, $\left\vert
1\right\rangle _{L}=|10\rangle $ (using two physical qubits per logical
qubit) is highly compatible with $H_{\mathrm{ex}}$, in the sense that
universal QC can be performed by controlling the \emph{single} parameter $%
J_{ij}^{x}$ in the Heisenberg ($J_{ij}^{x}=J_{ij}^{y}=J_{ij}^{z}$), XXZ ($%
J_{ij}^{x}=\pm J_{ij}^{y}\neq J_{ij}^{z}$), and XY ($J_{ij}^{x}=J_{ij}^{y}$, 
$J_{ij}^{z}=0$) instances of $H_{\mathrm{ex}}$, provided there is a Zeeman
splitting that distinguishes single-qubit $Z_{i}$ terms. This is done using
the \textquotedblleft encoded selective recoupling\textquotedblright\ method 
\cite{Lidar/Wu:01}. Furthermore, the $\{|01\rangle ,|10\rangle \}$ encoding
is a DFS for collective dephasing (where the bath couples only to system $%
Z_{2i-1}+Z_{2i}$ operators) \cite{Duan:98,Lidar:PRL99,Kempe:00}. A set of
logical operations on the code is $E=\{\overline{X}%
_{1}=(X_{1}X_{2}+Y_{1}Y_{2})/2$, $\overline{Y}_{1}=(X_{2}Y_{1}-Y_{2}X_{1})/2$%
, $\overline{Z}_{1}=(Z_{1}-Z_{2})/2\}$. Only the $\overline{X}_{1}$ term is
assumed to be directly controllable (by manipulation of $J_{12}^{x}$), while
the $\overline{Z}_{1}$ term can be turned on/off using recoupling \cite%
{Lidar/Wu:01}. The $\overline{Y}_{1}$ term can then be reached in a few
steps: 
\begin{equation*}
e^{-i\theta \overline{Y}_{1}}=e^{i(\pi /4)\overline{X}_{1}}e^{-i\theta 
\overline{Z}_{1}}e^{-i(\pi /4)\overline{X}_{1}}.
\end{equation*}%
The leakage errors are due to system-bath interactions where the system
terms include any of $X_{i},Y_{j},$ $X_{i}Z_{j}$ and $Y_{i}Z_{j}$, since as
is easily seen, such terms do not preserve the $\{|01\rangle ,|10\rangle \}$
code subspace. As pointed out first in \cite{Byrd/Lidar:ss}, the LEO can be
expressed as $(R_{L}^{\mathrm{E}(1)})_{2-\mathrm{DFS}}=\exp (i\pi \overline{X%
}_{1})=Z_{1}Z_{2}$, which means that it is implementable using just the
controllable $J_{12}^{x}$ parameter in the instances of $H_{\mathrm{ex}}$
mentioned above. This form for $(R_{L}^{\mathrm{E}(1)})_{2-\mathrm{DFS}}$ is
an instance of Eq.~(\ref{eq:R_L^E}). Note that, in agreement with our
general comments above, $(R_{L}^{\mathrm{E}(1)})_{2-\mathrm{DFS}}$ commutes
with every element of $E_{2-\mathrm{DFS}}$, meaning that logical operations
can be performed on the encoded subspace \emph{while eliminating leakage}.

\textit{Example 4}.--- The most economical encoding of a DFS against
collective decoherence requires three physical qubits per logical qubit. In
the tensor product of three qubits, there exists two doublets and a
quadruplet. In this case the logical qubit states are stored in the two
doublet states which represent the logical zero and one. Under the action of
collective errors, the two doublets mix within their respective subspaces,
but not with each other. The logical operations are formed from the
Heisenberg exchange interaction, which is known to be universal for this
code \cite{Kempe:00}. This example is given in detail in \cite{Wu/etal:02}
and \cite{Byrd/etal:04} where it is shown that it is possible to construct
efficient, canonical LEOs by using only the Heisenberg exchange interaction.
This follows in principle from the theorem in Ref.~\cite{Byrd/Lidar:ss} and
can be done in practice using the constructions in Refs.~\cite{Wu/etal:02}
and \cite{Byrd/etal:04}.


\subsection{Leakage Elimination on the 4-qubit DFS}

\label{sec:4qubitDFS}

The four-qubit DFS contains two singlet states for the representative qubit,
three triplets and a spin 2, or quintuplet. The singlet states represent the
logical zero and one of the DFS encoded qubit. It was shown in \cite%
{Kempe:00} that the exchange operation between the first and second qubits
in the computational basis will provide a logical $Z$ operation which we
denote $\bar{Z}$. However, on further analysis, we find that we cannot
create a \textquotedblleft canonical\textquotedblright\ LEO in the sense
described in Ref.~\cite{Wu/etal:02}. In this case, we require the more
general characterization of the LEO given above.

First we give an LEO that is appropriate in the Kempe, et al. \cite{Kempe:00}
basis. Let the (square of the) total angular momentum operator be denoted $%
\vec{S}^{2}$ with eigenvalue $S(S+1)$. Then 
\begin{equation*}
4\vec{S}^{2}=\left( \sum_{i}\vec{\sigma}_{i}\right) ^{2},
\end{equation*}%
where $\vec{\sigma}_{i}=(\sigma _{i}^{x},\sigma _{i}^{y},\sigma _{i}^{z})$
are the Pauli matrices acting on the $i$th qubit. Therefore 
\begin{equation*}
\frac{1}{2}S^{2}=\frac{1}{8}(12+2\sum_{i<j}\vec{\sigma}_{i}\cdot \vec{\sigma}%
_{j})
\end{equation*}%
gives an appropriate LEO of the form given in Eq.~(\ref{eq:R_Lmat}). This
can be seen as follows. On the $S=0$ (singlet) subspaces the operator gives
zero. On the $S=1$ (triplet) subspaces the operator gives $1$ and on the $S=2
$ (quintuplet) subspace it gives $3$. Therefore the appropriate LEO can be
obtained using 
\begin{equation*}
R_{L}=\exp (-i\pi S^{2}/2).
\end{equation*}%
Since the operator $S^{2}$ is composed of exchange interactions, it is also
experimentally available.


\section{Concluding Remarks}

We have provided an overview of the quantum error preventing strategies for
quantum computing devices. While the methods of QECC were motivated by error
correction methods for classical computing, the other methods are more
physically motivated. We believe the theory and practice of error prevention
in quantum computing systems is converging based upon strategies which
combine more than one of the error prevention techniques discussed here. The
progress is motivated by the desire to construct practical error prevention
schemes for near-future experiments. We hope that the overview given here
will provide a resource/review for novices/experts working in quantum
computing and that the progress concerning the elimination of leakage from
encoded spaces will aid in the development of practical error prevention
strategies.



\begin{thebibliography}{83}
\bibitem[{str()}]{strongnote}
\bibinfo{note}{One normally assumes that any reasonable candidate for a quantum
  computing device will satisfy the requirement that many gating operations can
  be performed within the decoherence time. Thus we somewhat reasonably assume
  external controls are strong compared to the strength of the interaction
  Hamiltonian so that the interaction Hamiltonian can be neglected during the
  gating operation.}
\end{thebibliography}


\end{document}